# Satellite quantum repeaters for a quantum Internet

Sundaraja Sitharama Iyengar,* and Mario Mastriani

*School of Computing & Information Sciences, Florida International University, 11200 S.W. 8th Street, Miami, FL 33199, USA*

## Abstract

This work presents a satellite alternative to quantum repeaters based on the terrestrial laid of optical fiber, where the latter have the following disadvantages: a propagation speed ($v$) equal to 2/3 of the speed of light ($c$), losses and an attenuation in the material that requires the installation of a repeater every 50 km, while satellite repeaters can cover greater distances at a speed $v = c$, with less attenuation and losses than in the case of optical fiber except for relative environmental aspects to the ground-sky link, i.e., clouds that can disrupt the distribution of entangled photons. Two configurations are presented, the first one of a satellite and the second one of two satellites in the event that both points on the ground cannot access the same satellite. Finally, a series of implementations for evaluating the performance and robustness of both configurations are implemented on a 5 qubits IBM Q processor.

ORCID Id:

S.S. Iyengar: 0000-0003-3203-833X

M. Mastriani: 0000-0002-5627-3935

---

*Electronic address: iyengar@cis.fiu.edu

# I. INTRODUCTION

Quantum entanglement [1-3] is the cornerstone within the modern discipline known as quantum communications [4-7], which is constituted by its most conspicuous representatives: quantum teleportation [8-12], quantum Internet [13-18], and quantum cryptography [19-21], being quantum key distribution (QKD) [22-24] the most famous tool of the latter.

Specifically, quantum Internet is emerging as the new version of the network of networks, which will provide the security factor as its most outstanding feature compared to the traditional or classic Internet. Two alternatives are presented to be able to get this super-net to all corners of the planet and beyond: by means of land lines of thousands of kilometers of optical fiber or via satellite links. In the first case, and given the characteristics of the materials with which the optical fibers are built, where light spreads at approximately two-thirds of the speed of light with losses and attenuations, it is necessary to use quantum repeaters [24-27] every 50 km. These repeaters are implemented in most cases by a technique known as entanglement swapping [28-34], which essentially consists of a game of extension of the entanglement range through the successive coupling of repeaters working by transitivity [24-27]. The second case is almost essential when the characteristics of the terrain are such that the laid of the optical fiber is practically impossible, then we must resort to satellites, which must perform some or all of the following functions: generation and distribution (to earth or another satellite) of entangled photon pairs, provisioning of the classic (generally electromagnetic) channel to convey the classic bits of disambiguation necessary to complete the processes related to the reconstruction of teleported states, and teleportation *per se*. In the two mentioned cases, i.e., through optical fiber or satellite links, we will need quantum memories [35] as buffers, in order to compensate for the arrhythmias of different times of trajectory of the entangled photons through paths of different lengths or through mediums with different speeds of light propagation.

On the other hand, although modern classical telecommunication relies on optical fibers, the direct transmission of photons through fibers is not practical for quantum communication over global distances because losses are too high [36]. The best available fibers have a loss of 0.15 dB/km at the optimal wavelength. This means, for example, that the time to distribute one entangled photon pair over 2000 km with a 1 GHz source exceeds the age of the universe [36]. For this reason, the current trend in quantum repeaters for quantum Internet is mainly focused on satellite links [36-38].

Although the aforementioned trend has been established, to date there is not a work which analyzes these satellite links by means of implementations on the currently available IBM Q platform [39], in particular, on its quantum processing units (QPU), because the quantum teleportation protocol uses the *if-then-else* statement which cannot be implemented in any current QPU based on superconductors [39]. Therefore, an uncovered space is generated in this area relative to the teleportation protocol [8] implemented on this type of platforms for quantum satellite repeaters. This is the main reason of this work, where an alternative but perfectly implementable and faithful teleportation protocol on any QPU is presented, which will allow us to study different configurations of satellite quantum repeaters at a significantly lower cost than if it were implemented on optical circuits [40].

Showing up next, the main prolegomena necessary to understand the outcomes obtained from a QPU are explained in Sec. II. Satellite quantum repeaters for one and two satellites are implemented on a five qubits IBM QPU in Sec. III. Finally, Sec. IV provides a conclusion and future work proposals.

## II. PROLEGOMENA

### A. Setup

As we can see in Fig. 1, Bloch's sphere has two poles called Computational Basis States (CBS), which can be expressed of several forms, yields,

$$North\ pole = Spin\ up = |0\rangle = \begin{bmatrix} 1 \\ 0 \end{bmatrix} \tag{1}$$

$$South\ pole = Spin\ down = |1\rangle = \begin{bmatrix} 0 \\ 1 \end{bmatrix} \tag{2}$$

All pure state can be represented on the Bloch's sphere [41-43] of Fig. 1 as a superposition of both CBS $\{|0\rangle, |1\rangle\}$, resulting in a wave-function like the following,

$$|\psi\rangle = \alpha|0\rangle + \beta|1\rangle = \alpha\begin{bmatrix} 1 \\ 0 \end{bmatrix} + \beta\begin{bmatrix} 0 \\ 1 \end{bmatrix} = \begin{bmatrix} \alpha \\ \beta \end{bmatrix} \tag{3}$$

where $|\alpha|^2 + |\beta|^2 = 1$, such that $\alpha \wedge \beta \in \mathbb{C}$ of a Hilbert's space [41]. Strictly, the complete wave-function will be

$$|\psi\rangle = e^{i\gamma}\left(cos\frac{\theta}{2}|0\rangle + e^{i\phi} sin\frac{\theta}{2}|1\rangle\right) = e^{i\gamma}\left(cos\frac{\theta}{2}|0\rangle + (cos\phi + i\,sin\phi) sin\frac{\theta}{2}|1\rangle\right) \quad (4)$$

where $0 \leq \theta \leq \pi$, $0 \leq \phi < 2\pi$ [41]. However, we can ignore the factor $e^{i\gamma}$ of Eq.(4), because it has no observable effects [44], and for that reason we can effectively write

$$|\psi\rangle = cos\frac{\theta}{2}|0\rangle + e^{i\phi} sin\frac{\theta}{2}|1\rangle, \quad (5)$$

with $\alpha = cos(\theta/2)$ and $\beta = e^{i\phi} sin(\theta/2)$, being then equal to the Eq.(4). The numbers $\theta$ and $\phi$, along with the radius of the sphere $r$, define a point on the unit three-dimensional sphere, as shown in Fig. 1, with Cartesian,

$$|\psi\rangle \equiv \begin{bmatrix} z = cos(\theta/2) = \alpha \\ x = cos\phi\,sin(\theta/2) = \beta_x \\ y = i\,sin\phi\,sin(\theta/2) = \beta_y \end{bmatrix} \quad (6)$$

or polar,

$$|\psi\rangle \equiv \begin{bmatrix} r \\ \theta \\ \phi \end{bmatrix} \quad (7)$$

representation, being $r = 1$. This level of detail in the analysis of the qubits on the Bloch sphere will be of paramount importance when analyzing the outcomes delivered by the different platforms on which the protocols will be implemented. Then, given a generic qubit on Hilbert space $H_{2\times 1}$ as that of Eq.(3), its density matrix in $H_{2\times 2}$ will be,

$$\rho_{|\psi\rangle} = |\psi\rangle\langle\psi^*| = \begin{bmatrix} \alpha \\ \beta \end{bmatrix}\begin{bmatrix} \alpha^* & \beta^* \end{bmatrix} = \begin{bmatrix} |\alpha|^2 & \alpha\beta^* \\ \beta\alpha^* & |\beta|^2 \end{bmatrix}, \quad (8)$$

where (•)* means complex conjugate of (•). On the other hand, the elements on the main diagonal of the density matrix of Eq.(8) will represent the probabilities or outcomes obtained as a consequence of the quantum measurement process [45, 46] of qubits, and which can also be obtained, taking into account Eqs.(1, 2, and 3), using the following procedure:

Probability of $|0\rangle$ $(\text{Po}|0\rangle) = \langle \psi^*|0\rangle\langle 0|\psi\rangle = |\alpha|^2 = \rho_{|\psi\rangle(1,1)}$ (9a)

Probability of $|1\rangle$ $(\text{Po}|1\rangle) = \langle \psi^*|1\rangle\langle 1|\psi\rangle = |\beta|^2 = \rho_{|\psi\rangle(2,2)}$ (9b)

Therefore, given a qubit on the Bloch sphere, in theory, we have the following elements to represent it:

- wave-function
- state-vector
- density matrix,
- Probabilities.

However, as we will see in the next subsection, in a simulator all the mentioned elements are generally available, while in the case of a QPU, only the probabilities of Eq.(9) are available.

**B. Implementation on quantum platforms**

From now on, we will use an interesting example, which consists in the generation of a simple qubit from a combination of gates of the type:

$$THTH|0\rangle = \begin{bmatrix} 1 & 0 \\ 0 & exp(\frac{i\pi}{4}) \end{bmatrix} \frac{1}{\sqrt{2}}\begin{bmatrix} 1 & 1 \\ 1 & -1 \end{bmatrix}\begin{bmatrix} 1 & 0 \\ 0 & exp(\frac{i\pi}{4}) \end{bmatrix} \frac{1}{\sqrt{2}}\begin{bmatrix} 1 & 1 \\ 1 & -1 \end{bmatrix}\begin{bmatrix} 1 \\ 0 \end{bmatrix}.$$ (10)

This qubit has the following characteristics:

*wave-function:*

$$|\psi\rangle = \begin{bmatrix} 0.9241 \\ 0.2706 + 0.2706\,j \end{bmatrix},$$ (11)

*state-vector:*

$$|\psi\rangle = \begin{bmatrix} 0.854 + 0.354\,j \\ 0.354 - 0.146\,j \end{bmatrix},$$ (12)

*density matrix:*

$$\rho_{|\psi\rangle} = \begin{bmatrix} 0.9241 \\ 0.2706 + 0.2706\,j \end{bmatrix}\begin{bmatrix} 0.9241 & 0.2706 - 0.2706\,j \end{bmatrix} = \begin{bmatrix} 0.854 & 0.25 + i0.25 \\ 0.25 - i0.25 & 0.146 \end{bmatrix},$$ (13)

*Probability of |0>:*

Po|0> = 85.35534 %, and

*Probability of |1>:*

Po|1> = 14.64466 %

For a more graphic description of what we have seen so far, we will turn to Quirk simulator [47], as it is an especially visual platform, which incorporates all the mentioned features, i.e.: Bloch's sphere (BS), density matrix (DM), and Chance of being ON if measured (Po|1>), among others. According to Fig. 2, for the qubit of Eq.(10) Quirk returns the following outcomes:

- *Bloch sphere (BS) representation of local state:*

  $r$ = +1.0000, $\varphi$ = -45.00º, $\theta$ = +45.00º

  $x$ = +0.5000, $y$ = -0,5000, $z$ = +0.7071

- *Density matrix (DM):*

  00: Probability of |0> (decimal 0) = 85.3553 %

  01: Coupling of |0> to <1| (decimal 0 to 1) = +0.250000+0.250000i

  10: Coupling of |1> to <0| (decimal 1 to 0) = +0.250000-0.250000i

  11: Probability of |1> (decimal 1) = 14.6447 %

- *Chance of being ON if measured (Po|1>):*

  Po|1> = 14.64466 %

Figure 2 shows us the density matrix (DM) of Eq.(10) qubit as follows: the radius of the green circle in every quadrant represents the modulus of the correspondent element, the orientation of the black segment means the angle (or phase) of the correspondent element, and the level of green in the elements whose positions are 00 and 11 means: Probability of |0> (Po|0>), and Probability of |1> (Po|1>), respectively. Moreover, Probability of |1> (Po|1>) has its own icon in Quirk [47], which allows us to automatically deduce the Probability of |0> (Po|0>) by means of a simple subtraction: Po|0> = 1 - Po|1>, which is absolutely concomitant with Eq.(10).

On the other hand, a conspicuous exponent in the world of quantum platforms is represented by IBM Q [39], which has the largest number of environments for the implementation of quantum circuits (in reality, environments for the expression of outcomes), in fact, four:

- *circuit composer*, which gives us information about:
    - number of classical and quantum registers,
    - qasm code (quantum assembly language), particularly, openqasm 2.0,
    - density matrix
    - state-vector with output state and phase angle $\theta\, e^{j\theta}$ in color, with *Y* axis (component amplitudes) and *X* axis (computational basis states), and
    - Measurement probabilities.
- *simulator* with a selector of the number of shots: 1, 1024, 4096, 8192. It can increase the number of shots to improve statistical accuracy. Its designation is: ibmq_qasm_simulator (Up to 32 qubits), being fairshare its RUN Mode.
- *qiskit notebooks (quantum information science kit)*, which supports Python 3.5, and the
- *processors (QPUs)*, which depend on the number of qubits available and according to their availability may have the following designations for:
    - *1 qubit*: ibmq_armonk
    - *5 qubits*: ibmq_london, ibmq_burlington, ibmq_essex, ibmq_ourense, ibmq_vigo, ibmq_rome, and ibmq_5_yorktown - ibmqx2,
    - *15 qubits*: ibmq_16_melbourne.

  Their RUN Mode is fairshare. All also with a selector of the number of shots: 1, 1024, 4096, 8192.

Figure 3 shows the qubit of Eq.(10) generated from ground state with its respective metrics on IBM Q [39] circuit composer, where:

a) it represents the quantum circuit, very similar to that of Fig. 2,

b) the height of the bar is the complex modulus of the wave-function, in this case is 0.9241 for |0>, and 0.3826 for |1>,

c) it is the real part of the state, with 85.3553 % for |0>, and 14.6447 % for |1>,

d) it is the real part density matrix of the state, and

$$\begin{bmatrix} 0.854 & 0.25 \\ 0.25 & 0.146 \end{bmatrix}$$

e) it is the imaginary part density matrix of the state.

$$\begin{bmatrix} 0 & 0.25 \\ -0.25 & 0 \end{bmatrix}$$

Finally, Fig. 4 represents the same qubit of Eq.(10) but on the other environments, where:

a) they are the Probabilities (outcomes) of IBM QPU simulator [39] for 1024 shots, with 85,449 % for |0>, and 14,551 % for |1>,

b) they are the Probabilities (outcomes) of the IBM QPU [39] for ibmq_armonk processor for 1024 shots too, with 83,789 % for |0>, and 16,211 % for |1>, and

c) they represent the Run details of IBM QPU [39] for ibmq_armonk processor of 1 qubit.

There is an absolute coincidence between the outcomes obtained with the Quirk platform [47] of Fig. 2 with those from the IBM Q [39] circuit composer and simulator in Figures 3 and 4. Instead, the difference is evident between the outcomes of the simulator in Fig. 4 (a) and the QPU (ibm_armonk) in Fig. 4 (b). This difference between simulator and processor will become more noticeable in the case of the teleportation protocols implemented in later sections as part of the satellite configurations used, which will be mainly due to the fact that said protocols will involve many more gates than those used in Fig. 3 (a) to generate the qubit of Eq.(10), since a greater number of gates introduce a greater amount of noise, and greater dependence on the decoherence associated with those gates [45, 46].

### C. Safe vs unsafe quantum teleportation protocols

Figure 5 (a) represents the traditional quantum teleportation protocol on IBM Q [39] with several clearly defined sectors, by columns: generation of the Einstein-Podolsky-Rosen (EPR) pair (i.e., the entangled pair), qubit preparation (in this particular, the same qubit of Eq.10), the Bell State Measurement (BSM) module; and by rows: Alice's side above the green line, and Bob's side below it.

The problem in implementing the quantum teleportation protocol of Fig. 5 (a) lies in the inability of QPUs based on superconductor to incorporate the *if-then-else* statement into their arsenals, among others elements like qubit reset gate, definition *per se* of new gates, and use of quantum measurements in intermediate instances of a quantum circuit. Therefore, this protocol can only be implemented in the

following environments:

- circuit composer (almost unlimited in the number of qubits),

- qiskit notebooks, and

- Simulator: ibmq_qasm_simulator (up to 32 qubits)

Although simulators are a powerful and very important working tool within quantum technology, the reality is that having exclusively their outcomes takes us away from practical reality, in fact, from reality that can be implemented on optical circuits [40], which constitute the putative environment in relation to the implementation of satellite quantum repeaters [36-38]. For this reason, we will resort to an alternative version to the protocol of Fig. 5 (a) and that we can see in Fig. 5 (b) also on IBM Q [39]. This version teleports both qubits and continuous variables with the same outcomes as the original version but does not use quantum measurements (or single photon detectors: SPD) or the *if-then-else* statement inside the BSM module. However, while we will call the original version safe, we will call the alternative version unsafe. The reason is simple and explained in the *pros* and *cons* of the proposed version (i.e., the unsafe one):

- *Pros:*
    - this version gives exactly the same outcomes as the original version, allowing to faithfully test the concept (or feasibility) for satellites configurations of the next section on a QPU of IBM Q [39] with free access via the web, and
    - this version allows establishing a faithful parallel with the versions implemented on optical circuits [40] at a much lower cost, given that while an optical laboratory costs hundreds of thousands (and even millions) of dollars, the access to QPUs of IBM Q [39] is totally free and via the web.
- *Cons:*
    - this version is not safe in practice because, as will be seen shortly, a manifestation of what we want to teleport must be transmitted through the classic channel, in other words, a hacker could recover the qubit that is being attempted to teleport through a simple unitary transfor-

mation. This would force the classic channel to be encrypted in a special way, losing all meaning the quantum teleportation itself, and

- this version forces us to send more information through the classical channel than through the quantum channel, i.e., more information than that which we want to teleport, which is absurd, so, and again we ask ourselves, what is the sense of using teleportation then?

Let us see in detail the *cons*. Figures 5 (a) and (b) show us an absolute coincidence of both protocols until time $t_6$, the difference begins precisely there, in fact, there will be 3 times of particular importance in this analysis: $t_3$, $t_5$, and $t_6$, which can be seen in detail in Fig. 5 (c) on the Quirk [47] platform thanks to its conspicuous metrics: DM, Po|1>, and BS. It is precisely Fig. 5 (c) that allows us to complete Table I, without a doubt, since $t_0$ to $t_6$, which confirms the complete coincidence of both protocols up to $t_6$.

The right margin of Figs. 5 (a) allows us to access the result of measurements made using single photon detectors (SPD), obtained at times $t_7$ for q[0], and $t_8$ for q[1] in the traditional protocol [9, 10]. These outcomes, two classical bits (i.e., 0 or 1), represent the Bell base that arises randomly from the quantum measurements of the BSM, i.e., any of the following four:

$$|\beta_{00}\rangle = |\Phi^+\rangle = \frac{1}{\sqrt{2}}(|00\rangle + |11\rangle), \quad |\beta_{10}\rangle = |\Phi^-\rangle = \frac{1}{\sqrt{2}}(|00\rangle - |11\rangle),$$
$$|\beta_{01}\rangle = |\Psi^+\rangle = \frac{1}{\sqrt{2}}(|01\rangle + |10\rangle), \quad |\beta_{11}\rangle = |\Psi^-\rangle = \frac{1}{\sqrt{2}}(|01\rangle - |10\rangle).$$
(14)

In the traditional version [8], the two mentioned classic bits, of possible values 0 or 1, are those that travel through the classic channel from Alice to Bob in order for Bob to apply the corresponding unitary transformation (according to both bits) for the recovery of the teleported state.

Instead, in the case of the right margin of Fig. 5 (b) for the unsafe protocol, what we have in $t_6$ for q[0] and q[1] is what Fig. 5 (c) indicates in $t_6$ and the Table I in its last row. In other words, what travels through the classic channel for unsafe protocol is much more information than traditional one, yields:

$$\{q[0] = H(CNOT(|\psi\rangle \otimes |\beta_{00}\rangle)), q[1] = |\beta_{00}\rangle\}$$ for the unsafe version, vs

{0/1, 0/1} for the traditional version.

As if the previous problem was not enough, applying the following sequence of gates to q[0], a hacker would obtain: $\text{CNOT}(\text{H}(q[0])) = |\psi\rangle \otimes |\beta_{00}\rangle$, i.e., the state to be teleported $|\psi\rangle$ projected on the Bell's basis $|\beta_{00}\rangle$, hence the reason for the name unsafe, whereas in the traditional protocol [8], a hacker could never reconstruct the qubit to be teleported having only the classic bits.

Summing-up, although the unsafe version is not safe or practical for satellite quantum repeaters [36-38], it allows to faithfully establish the functionality of the process and thus study the feasibility of the different configurations at a considerably lower cost than if it were a laboratory experiment on a circuits optical [40]. Therefore, from now on, all the results will be obtained using this protocol.

## III. SATELLITE QUANTUM REPEATERS

This section presents two satellite configurations of quantum repeaters [36-38], one and two satellites, for which we will use both the Quirk simulator [47] and the IBM Q family simulators and QPUs [39]. In this sense, we must remember that the focus of this work is exclusively protocol-oriented and not the satellite technology itself, i.e., we do not clarify technological details of: the satellite platforms themselves, the links through telescopes, characteristics of said telescopes, tracking beam systems, quantum memories, among many other technological tips.

On the other hand, while in satellite quantum repeaters we will work exclusively with entangled photon pairs, in IBM Q QPUs we will work with entangled states thanks to superconductor technology [39]. Besides, on the Quirk simulator [47] we will implement the original protocol [8] (or safe) while on the IBM Q platforms we will implement the proposed protocol (or unsafe), in this way we can check the equality between the outcomes of both versions on two platforms completely different.

In this paper, in the case of two or more satellites, we will consider repeater from the second satellite (inclusive) forward, while in the case of a single satellite we speak exclusively of satellite link. Moreover, we will use a single satellite when both Alice and Bob can access it without difficulties associated with the terrain or climate, in particular clouds that prevent the distribution of the entangled photons. Instead, we will use two satellites when some of the ends of the communication (i.e., Alice or Bob) cannot access the same satellite. In the latter case, one of the satellites will act as the master and the other as the slave, where the slave only distributes entangled photons and performs an electromagnetic communication link to transmit the disambiguation bits, while the master does this and also carries an on-board Bell State Measurement (BSM) module.

### A. One satellite

We present here a single satellite configuration for both the Quirk simulator [47] and the IBM Q platforms [39], starting with Quirk.

Figure 6 presents the original quantum teleportation protocol (safe) on Quirk [47] between Alice (red) and Bob (blue) thanks to only one satellite (green). Quantum circuit highlighted in red (Alice) is on land, which is constituted by a qubit generation module in this case, via this particular combination of gates: THTH|0> of Eq.(10); and the Bell State Measurement (BSM) module formed by a CNOT and a H (Hadamard) gates with a couple of quantum measurement gates [35, 36] representing the Single Photon Detectors (SPD), which in practice are four, one associated with each Bell's base [1-3, 41-43]. Quantum satellite, highlighted in green, which is inserted interstitially between the terrestrial modules in red (Alice) and blue (Bob), emits entangled photons to the Earth (to Alice and Bob points) and constitutes the classic channel through which Alice sends Bob the result of the measurement made in the BSM. This result are represented by two classic disambiguation bits, which represent the Bell's base that emerged randomly in the mentioned measurement. Quantum circuit highlighted in blue (Bob) is on land, which represents the unitary transforms (based on Pauli's matrices) which Bob must apply to reconstruct the teleported state. The total coincidence between the Quirk's metrics on Alice's side (i.e., q[0]) as well as Bob's side (i.e., q[2]) demonstrates the success of the teleportation.

Figure 7 allows us to see a practical representation of the regrouping of Figure 6 with a clear correspondence between all the involved blocks. All begins on point A (on land), where the block labeled as BSM is inside macro-block highlighted in red in Figure 6. The point B (on land) receives an element of the EPR pair (orange rays) and the classical bits of disambiguation (gray rays) emitted by the quantum satellite with which Bob applies a unitary transformation U (inside a block highlighted in blue in Fig. 6) and thus reconstructs the teleported state. Therefore, in this case the satellite performs a simple but efficient relay of hybrid communications (quantum and classic) between points A and B.

Figure 8 shows the proposed protocol on the circuit composer of IBM Q [39] for the qubit of Eq.(10). We can observe three sectors very good defined: the qubit generation module to be teleported, the qubit teleporter block with the modified Bell State Measurement (BSM) inside it, and the quantum measurement module. The modified BSM module does use neither quantum measurement gates (i.e., single-photon detectors) nor *if-then-else* statements.

The metrics in Fig. 9 show that the qubit to be teleported from point (1) in Fig. 8, and whose state we know in full from Section II.B, is identical to that finally teleported and present in the point (2) of Fig. 8, however, if we look in detail at Fig. 5 on $t_6$, as well as the last row of Table I, we will see that the states of the three qubits are different, so the No-Cloning Theorem [48] is never violated.

In detail, the metrics of Fig. 9 for the point (2) of Fig. 8 gives a state-vector of 0.462 from 000 to 011, and 0.135-0.135j from 100 to 111. Besides:

a) It is the complex modulus of the wave-function will have:

Amplitude: 0.462, Phase angle: $2\pi$, from 000 to 011, and

Amplitude: 0.191, Phase angle: $7\pi/4$, from 100, to 111.

b) It is the real part of the state (histogram) with:

Po|0> = 85.35534 %, and

Po|1> = 14.64466 %.

c) It is the real part density matrix of the state, with:

*darker light blue tiles* = 0.21, and

*light blue tiles* = 0.06.

d) It is the imaginary part density matrix of the state, with:

*white tiles* = 0,

*light blue tiles* = 0.06, and

*pink tiles* = -0.06.

The results coincide exactly with those of Sec. II.B for qubit of Eq.(10), i.e., Figs. 2 and 3, which indicates that the teleportation was successful, however, the last metrics are associated directly with the IBM Q [39] circuit composer, which obviously has to do with a simulator and not with a physical machine. For this reason, the analysis that emerges from Fig. 10 is so important since it involves a QPU. In this case, the said figure shows us the same experiment with the following results:

a) Probabilities (outcomes) on the IBM QPU simulator [39] for 1024 shots, with:

Po|0> = 87.012 %, and

Po|1> = 12.988 %.

b) Probabilities (outcomes) for the ibmq_rome processor (5 qubits) of IBM QPU [39] for 1024 shots too, with:

Po|0> = 74.316 %, and

Po|1> = 25.684 %.

c) The Run details of ibmq_rome processor (5 qubits) of IBM QPU [39].

Where the RUN Mode was fairshare in both cases. Besides, it is evident the notorious difference between the outcomes of the simulator (a) and those of the selected QPU (b), for the same reasons explained in Sec. II.B, and as predicted at that time, the difference in the number of gates between the implementations of Figs. 8 and 3 implies a higher incidence of noise and decoherence in the case of teleportation compared to the single generation of the qubit, respectively.

### B. Two satellite

In this case, a configuration of two satellite quantum repeater is presented, where a satellite (called master) will have more task than the other one (called slave), although both will distribute entangled photons and will make electromagnetic communications links in order to transmit the classic bits of disambiguation necessaries for the teleportation protocols. As in the previous case, of a single satellite, here we will also carry out implementations on the Quirk simulator [47] and on a five qubits processor of the IBM Q [39] platform, starting again with Quirk.

Figure 11 presents the original quantum teleportation protocol (safe) on Quirk [47] between Alice (red) and Bob (blue) thanks to two satellites: slave or passive (green), and master or active (yellow). In this case and for geographical reasons, Alice does not see the yellow satellite, while Bob does not see the green satellite. The green satellite performs the same functions as in the experiment of Fig. 6, the difference is in the tasks performed by the yellow satellite, which has the responsibility of making an intermediate teleportation in order to ensure the definitive link between Alice and Bob. The complete match between the three sets of Quirk's metrics (in qubits: q[0], at the beginning of the red block, q[2], in the middle of the yellow block, and q[4], at the end of the blue block) ensures a successful final teleportation.

Figure 12 shows the complete satellite configuration, where for geographical and/or climatological reasons, Alice cannot see the yellow satellite (master), and Bob cannot see the green satellite (slave). The master satellite (green) distributes a pair of entangled photons, one to Alice (on land) and another one to the slave satellite (yellow), while the slave satellite distributes another entangled pair, where it conserves one photon and transmits to Bob the another one. The distribution of all entangled photons are represented by orange rays. The red BSM block on Alice's side transmits the first pair of classical bits of disambiguation resulting from the quantum measurement to the master satellite (green), while the latter transmits the second pair of classical bits of disambiguation resulting from another quantum measurement corresponding to an on-board BSM to the slave satellite (yellow). Finally, this satellite transmits the last pair of classical bits to Bob, which will use them to configure the unitary transformation U (blue) to obtain the final teleported state. All the classical links are represented with gray rays.

Figure 13 represents the experiment of Fig. 11 implemented on IBM Q [39]. We can distinguish five well-defined sectors separated by barriers, which can be grouped into four better defined operationally: the qubit to be teleported, a first qubit teleporter, a second qubit teleporter, and the quantum measurement module. In these last four sectors, we have overlapping elements from different devices of Fig. 12. In addition, we can see three points which will be analyzed below, in the following figures, where the metrics of Fig. 14 for the point (3) of Fig. 13 gives a state-vector of 0.231 from 00000 to 01111, and 0.068-0.068j from 10000 to 11111. Besides:

a) It is the complex modulus of the wave-function will have:

   Amplitude: 0.231, Phase angle: $2\pi$, from 00000 to 01111, and

   Amplitude: 0.096, Phase angle: $7\pi/4$, from 10000, to 11111.

b) It is the real part of the state (histogram) with:

Po|0⟩ = 85.35534 %, and

Po|1⟩ = 14.64466 %.

c) It is the real part density matrix of the state, with:

   *darker light blue tiles* = 0.05, and

   *light blue tiles* = 0.02.

d) It is the imaginary part density matrix of the state, with:

   *white tiles* = 0,

   *light blue tiles* = 0.02, and

   *pink tiles* = -0.02.

These results indicate the complete coincidence between the qubit prepared at point (1) and the qubit recovered at point (3), both in Fig. 13. Therefore, both must also coincide with the qubit at the intermediate point of the same figure, i.e., at point (2). However, these results come from the IBM Q circuit composer environment, therefore, it only remains to evaluate the results we will obtain on some of its QPUs.

Figure 15 shows us the same experiment with the following results:

d) Probabilities (outcomes) on the IBM QPU simulator [39] for 1024 shots, with:

   Po|0⟩ = 84.570 %, and

   Po|1⟩ = 15.430 %.

e) Probabilities (outcomes) for the ibmq_rome processor (5 qubits) of IBM QPU [39] for 1024 shots too, with:

   Po|0⟩ = 72.168 %, and

   Po|1⟩ = 27.832 %.

f) The Run details of ibmq_rome processor (5 qubits) of IBM QPU [39].

Where the RUN Mode was fairshare in both cases. Now, if we compare the results of Figs. 15 (b), for two teleportations, and 10 (b), for only one teleportation, it is evident that the increase (almost double) in the number of gates implies a greater divergence between the theoretical result and that obtained on the QPU, confirming what we have been saying in all the experiments in this section, i.e., there is a trade-off between the amount of satellite quantum repeaters and the noise associated with that amount, therefore, two options are hinted at as a result of this analysis:

a) to use satellites of higher orbit, i.e. farther from Earth, and thus use fewer satellites, so facilitating the line of sight between the satellite and the ends that need communicate, eventually compromised by the terrain (mountains, curvature of Earth, urbanization, etc), or

b) to explore a new satellite configuration that allows complete teleportation with fewer gates and the same number of satellites.

## IV. CONCLUSIONS AND FUTURE WORKS

### A. Conclusions

The experiments in Sec. III carried out on the IBM Q QPU [39] thanks to a modified version of the original quantum teleportation protocol [8], allow us to evaluate the viability, robustness (i.e., immunity to noise), as well as the most appropriate configurations depending on the geographic and climatic characteristics of the satellite quantum repeaters at zero cost. In other words, the advent of freely accessible IBM Q [39] type processors via the web allows us to create a laboratory of satellite quantum repeaters on our own personal computers at the cost of connecting to the Internet.

### B. Future works

Future lines of research in this area will necessarily involve protocols for bidirectional quantum teleportation based on satellite configurations, in order to contemplate the round trip in the flow of information without changing the tasks assigned to each actor, for example, although the communication is bi-directional, it will always be the satellite who distributes the pair of entangled photons.

The problems associated with entanglement swapping [28-34], especially those related to its implementations on land, i.e., its terrestrial laid of optical fiber, led to the development of this work. Therefore, given the encouraging results obtained in Sec. III with both a simulator and with a QPU, it is evident that this work has opened a new range of possibilities, giving rise to a new series of developments as alternative protocols even more efficient for satellite quantum repeaters [36-38] for quantum Internet [13-18].

**Data Availability**

The experimental data that support the findings of this study are available in ResearchGate with the identifier https://doi.org/10.13140/RG.2.2.28199.29600.

**Acknowledgements**

M. Mastriani thanks to Craig Gidney, head of Quirk, currently at Google, for his permanent willingness to answer all of our collaborators' questions about the correct interpretation of Quirk's tools, and for creating such a powerful, intuitively graphical, and pedagogical tool like Quirk. A special acknowledgment to Antonio D. Corcoles-Gonzalez, and Jay M. Gambetta of IBM Q, as well as to their community for their important clarifications about the practical implementation of the *if-then-else* statement on their Quantum Processing Units (QPUs).

**Competing interests**

Authors declare they has no competing interests.

**Author Contributions**

SSI is responsible for the project's conceptualization, and management, which concludes in this paper. The effort was planned and supervised by SSI. MM designed the study, conceived the protocols and the satellite configurations, designed the quantum circuits, performed the experiments, and wrote the first version of the paper. SSI analyzed the results. SSI reviewed the first version of the paper. SSI wrote the final version of the paper. SSI and MM read and approved the final manuscript.

**Funding**

This research received no external funding.

---

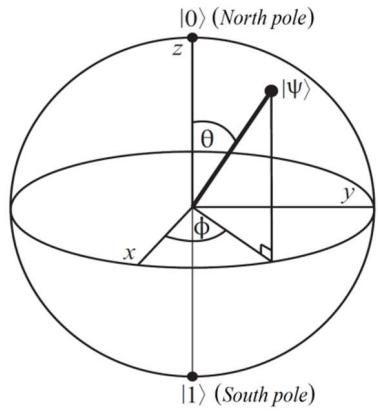

FIG. 1: Bloch's Sphere.

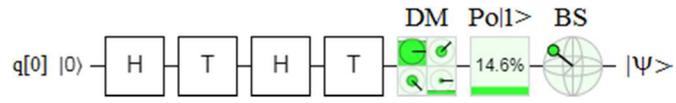

FIG. 2: Qubit of Eq.(10) generated from ground state with its respective metrics on Quirk [47]: density matrix (DM), Probability of |1> (Po|1>), and Bloch's sphere (BS).

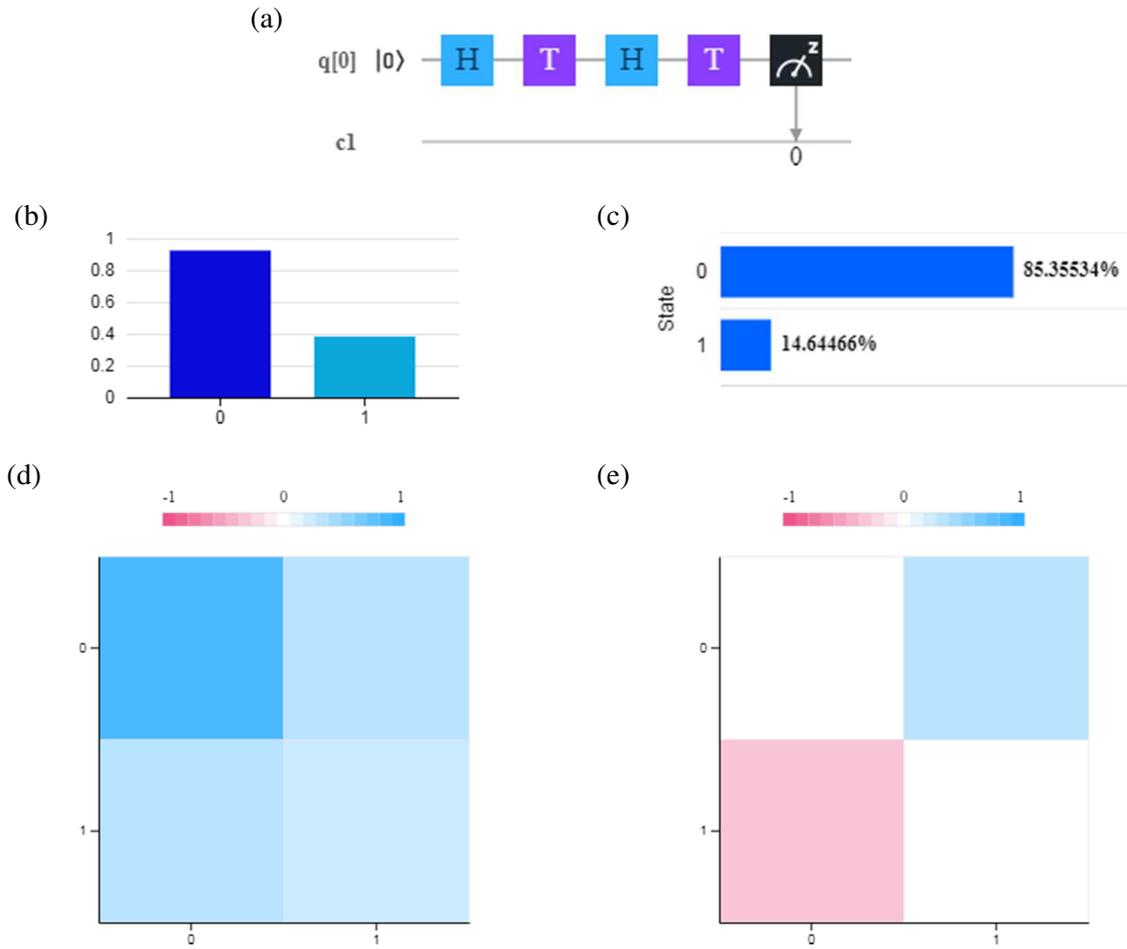

FIG. 3: Qubit of Eq.(10) generated from ground state with its respective metrics on IBM Q [39]: a) quantum circuit, b) height of the bar is the complex modulus of the wave-function, c) real part of the state, (d) real part density matrix of the state, and e) imaginary part density matrix of the state.

(a)

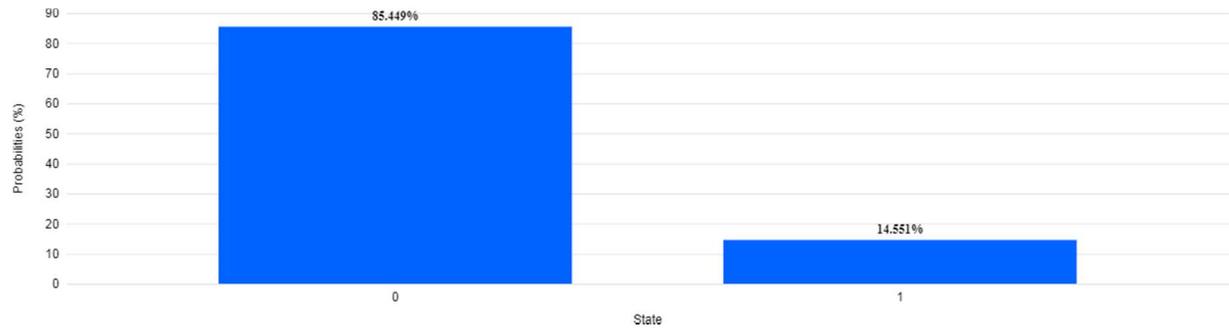

(b)

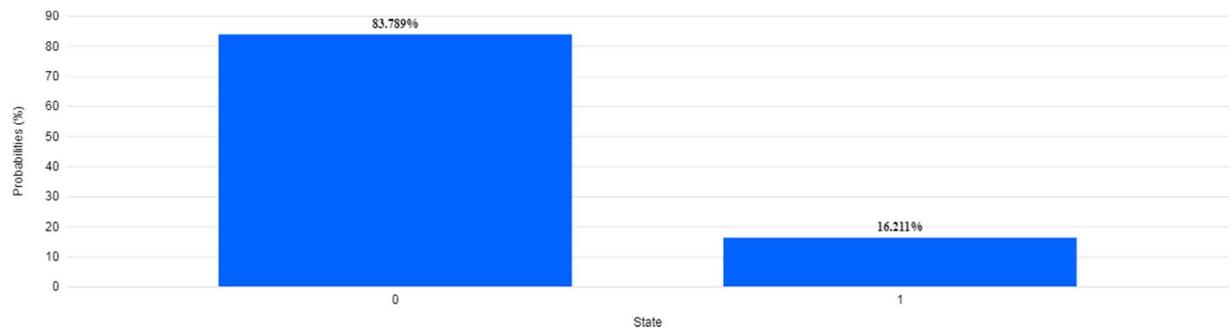

(c)

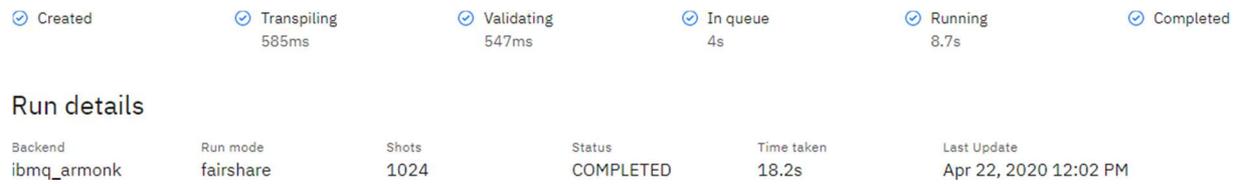

FIG. 4: Qubit of Eq.(10): a) Probabilities (outcomes) of IBM QPU simulator [39], b) Probabilities (outcomes) of IBM QPU [39] for ibmq_armonk processor, and c) Run details of IBM QPU [39] for ibmq_armonk processor.

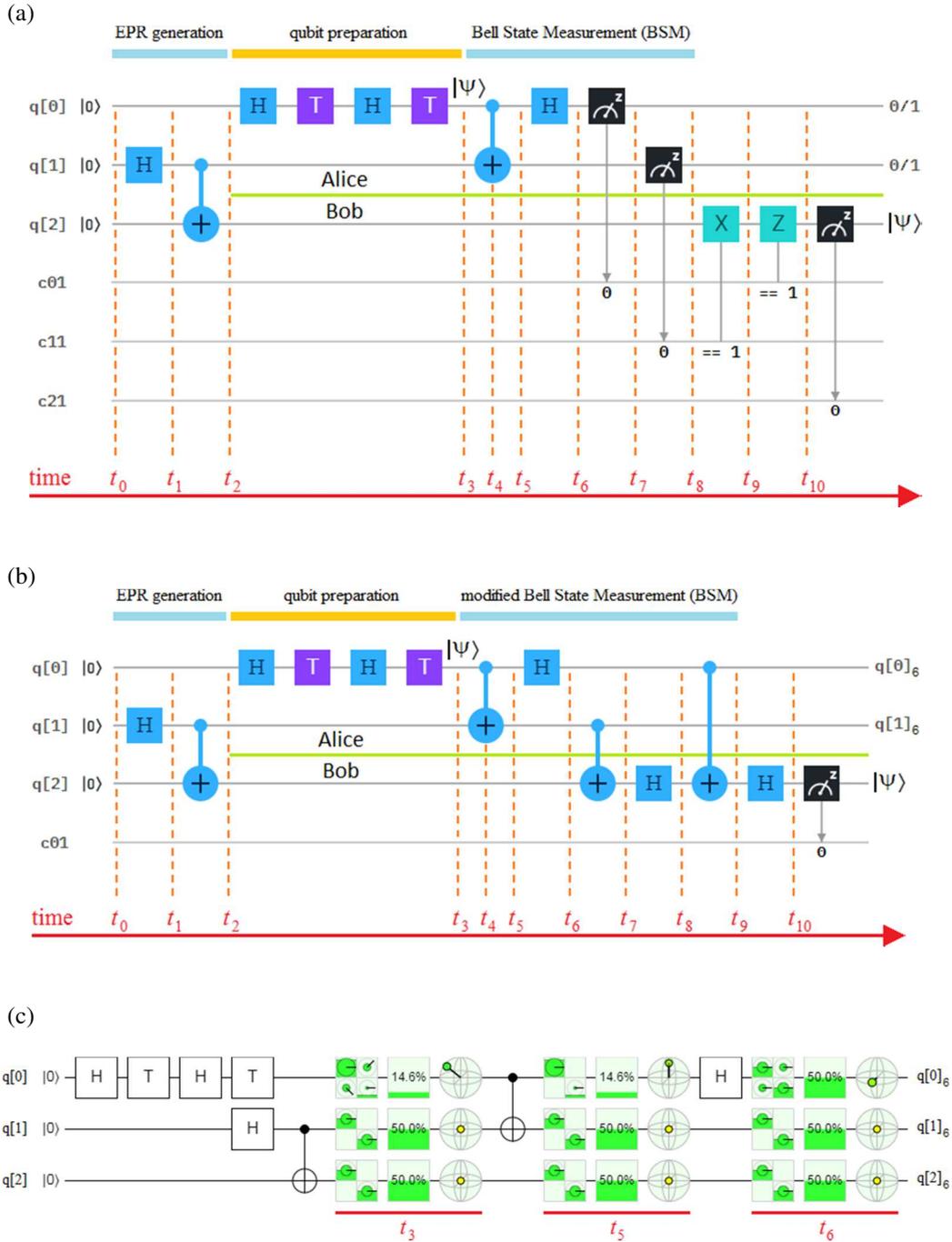

FIG. 5: Quantum teleportation protocols: a) traditional version on IBM Q [39], b) modified version on IBM Q [39], and c) the common sector between previous versions (a and b) since $t_0$ to $t_6$ on Quirk [47].

Table I. Protocols of Fig. 4 since $t_0$ to $t_6$.

| time | qubits | | |
| --- | --- | --- | --- |
| | Alice's side | | Bob's side |
| | q[0] | q[1] | q[2] |
| 0 | $\|0\rangle$ | $\|0\rangle$ | $\|0\rangle$ |
| 1 | $\|0\rangle$ | $\|+\rangle$ | $\|0\rangle$ |
| 2 | $\|0\rangle$ | $\|\beta_{00}\rangle$ | $\|\beta_{00}\rangle$ |
| 3 | $\|\psi\rangle$ | $\|\beta_{00}\rangle$ | $\|\beta_{00}\rangle$ |
| 4 | $q[0]_4 = \|\psi\rangle \otimes \|\beta_{00}\rangle$ | $\|\beta_{00}\rangle$ | $\|\beta_{00}\rangle$ |
| 5 | $q[0]_5 = \text{CNOT}(\|q[0]_4\rangle)$ | $\|\beta_{00}\rangle$ | $\|\beta_{00}\rangle$ |
| 6 | $q[0]_6 = H\|q[0]_5\rangle$ | $\|\beta_{00}\rangle$ | $\|\beta_{00}\rangle$ |

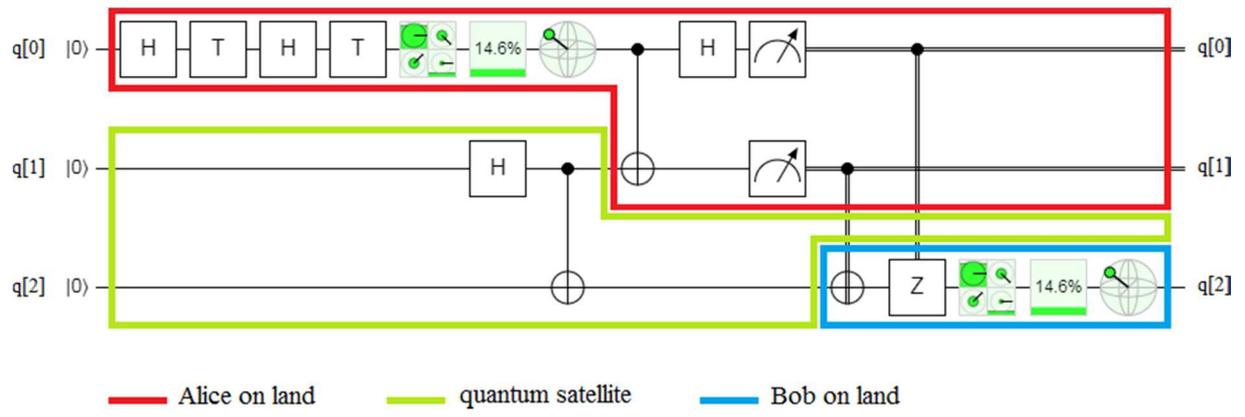

FIG. 6: Quantum teleportation protocol on Quirk [47] between Alice (red) and Bob (blue) thanks to only one satellite (green).

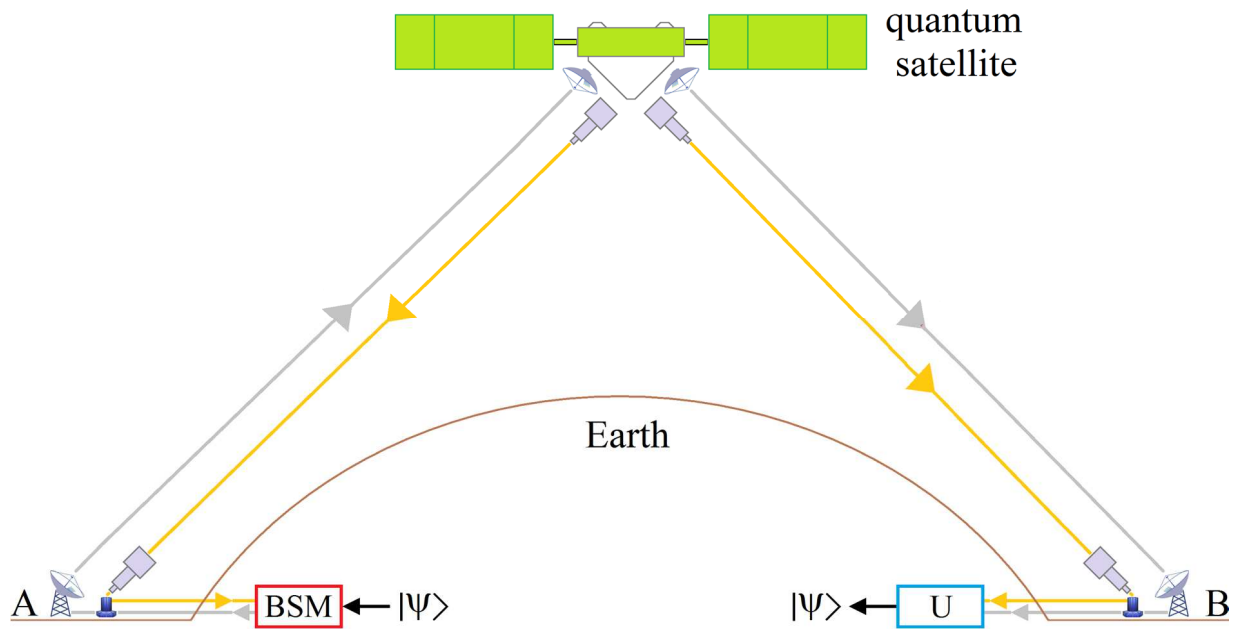

FIG. 7: One satellite to communicate Alice (A point) and Bob (B point), where the color code is respected with the Bell State Measurement (BSM) module on Alice's side in red, the unitary transform U on Bob's side in blue, and the satellite in green. Both Alice and Bob see the only satellite without difficulty. The orange rays represent the entangled photons scattered across the satellite, while the gray rays represent the electromagnetic link that drives the transmission of the classic bits of disambiguation that Bob needs to rebuild the teleported state.

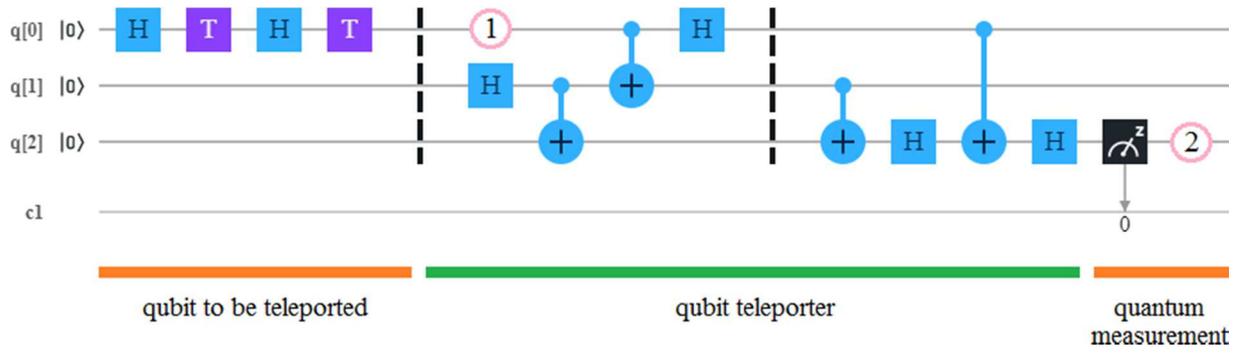

FIG. 8: The experiment of Fig. 6 implemented on IBM Q [39]. The qubit to be teleported is presented on point (1). After the qubit teleporter procedure, the qubit disappears from Alice's side and appears in point (2) on Bob's side.

(a) (b)

(c) (d)

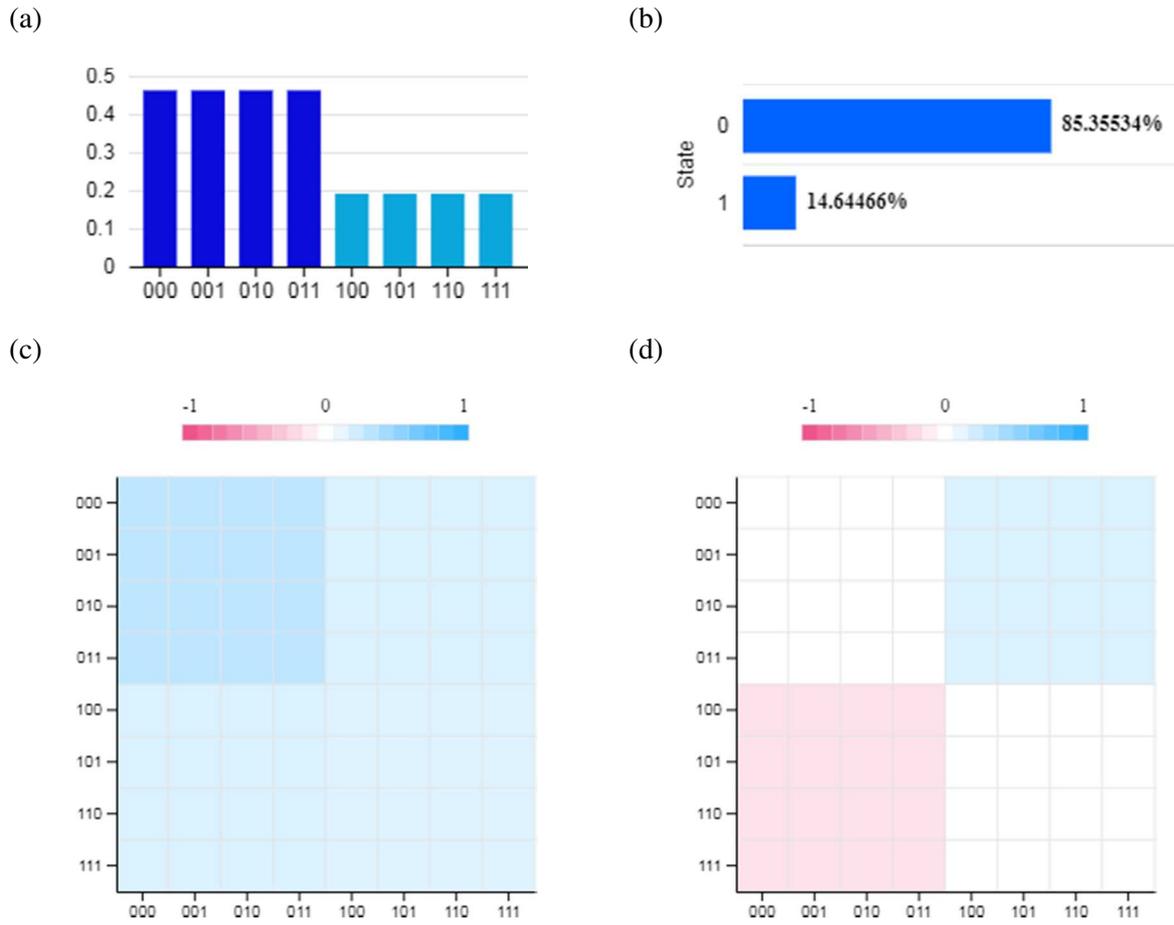

FIG. 9: Results of the teleportation of the state of Fig. 6 on IBM Q [39]: a) the height of the bar is the complex modulus of the wave-function, b) is the real part of the state (histogram), c) is the real part density matrix of the state, and d) is the imaginary part density matrix of the state.

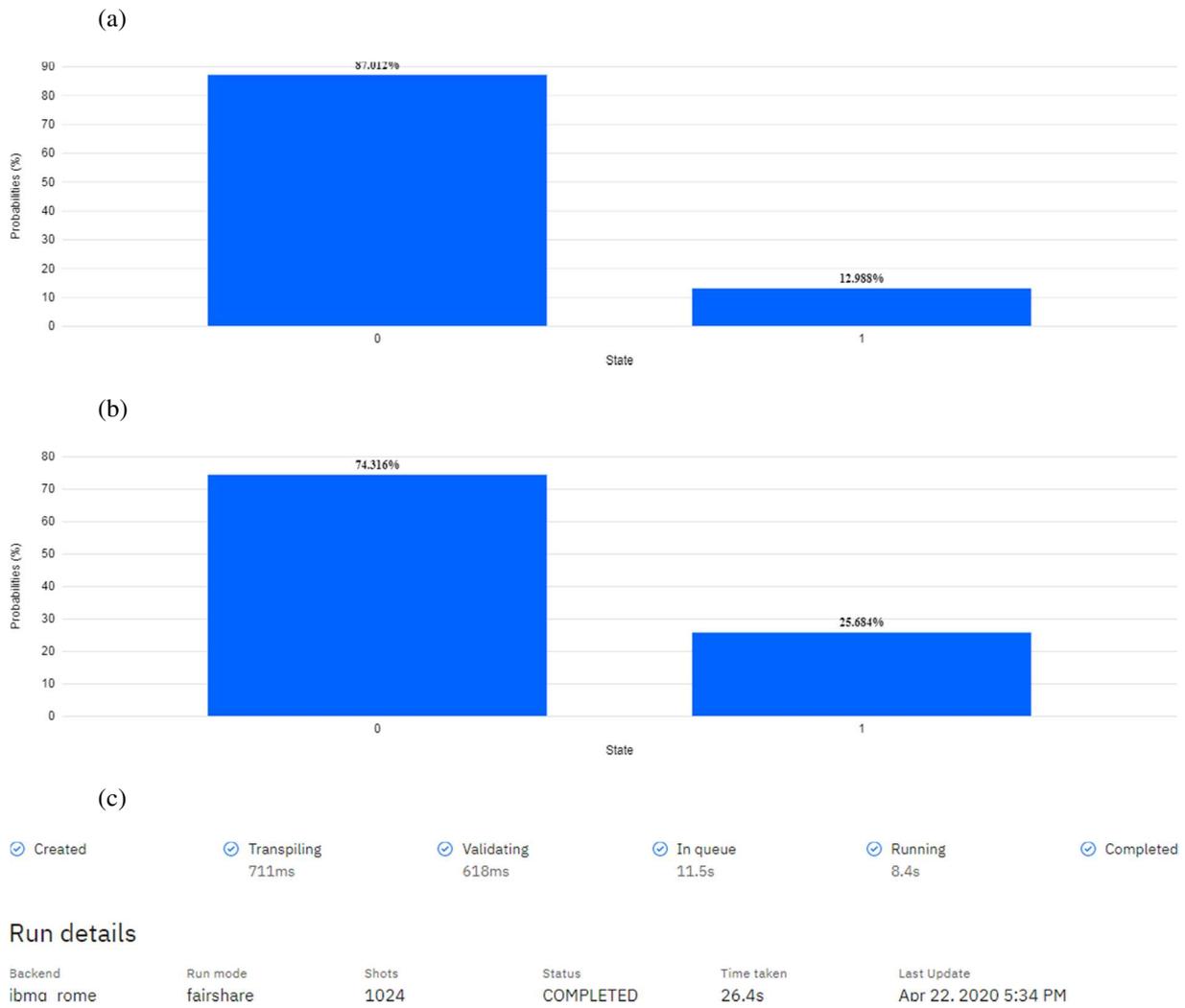

FIG. 10: Results of the teleportation of the state of Fig. 6 on IBM Q [39]: a) Probabilities (outcomes) of IBM QPU simulator [39], b) Probabilities (outcomes) of IBM QPU [39] for ibmq_rome processor, and c) Run details of IBM QPU [39] for ibmq_rome processor.

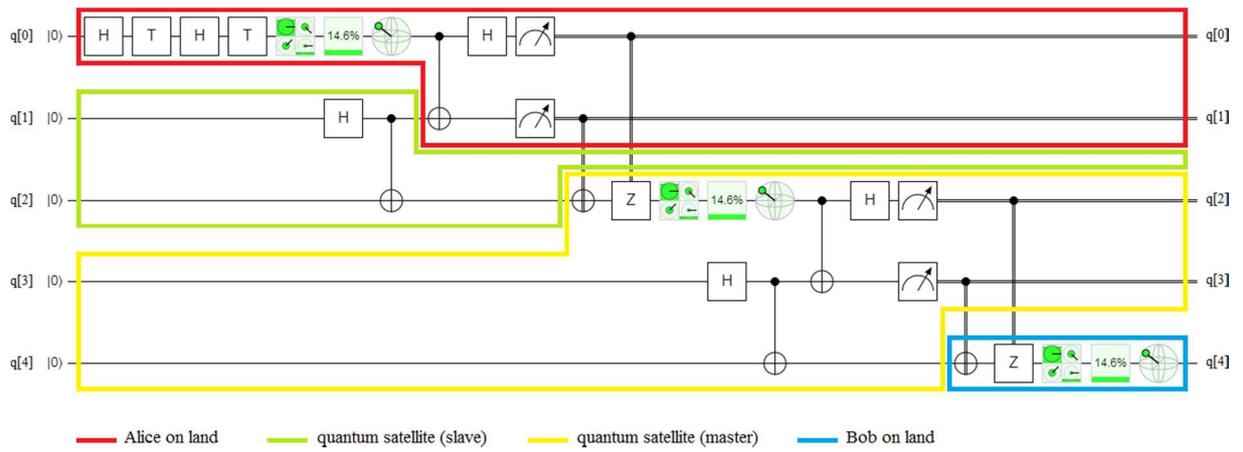

FIG. 11: Quantum teleportation on Quirk [47] between Alice (red) and Bob (blue) thanks to two satellites: slave or passive (green), and master or active (yellow).

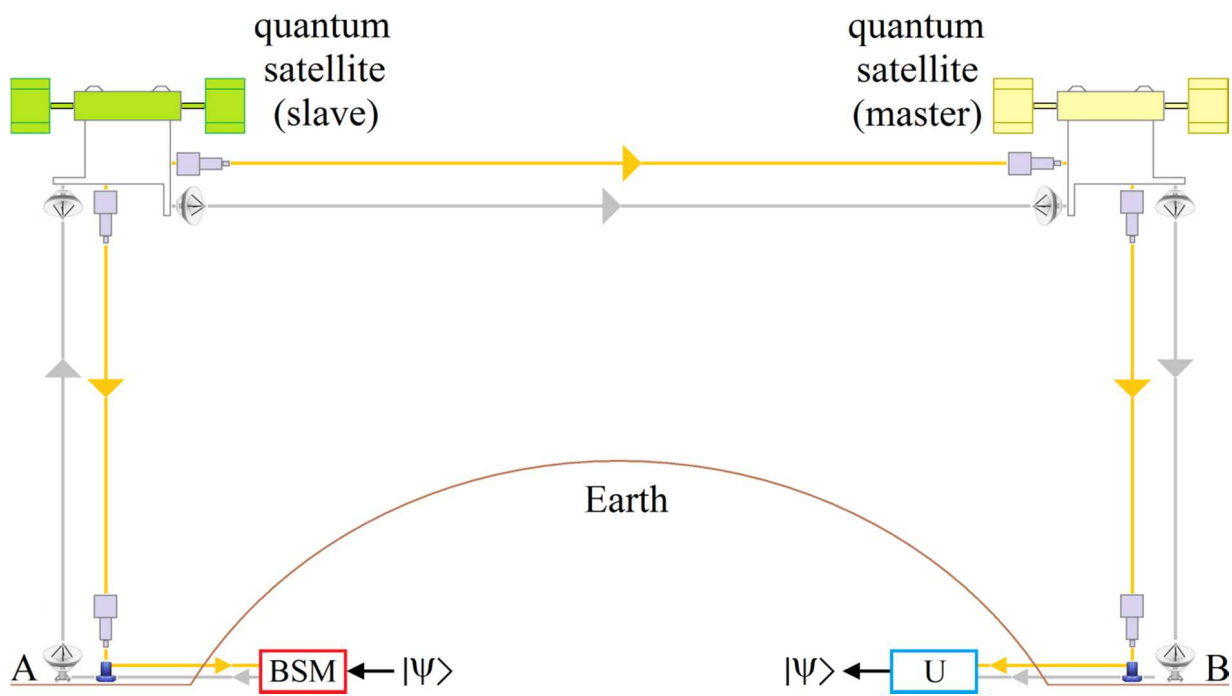

FIG. 12: Two satellites to communicate Alice (A point) and Bob (B point), where the color code is respected again with the Bell State Measurement (BSM) module on Alice's side in red, the unitary transform U on Bob's side in blue, the green satellite (slave only seen by Alice), and the yellow satellite (master only seen by Bob). The characteristics of this configuration require that the second satellite (master in yellow) have an additional function compared to the first one (slave in green) because an intermediate teleport is required to ensure the complete transmission.

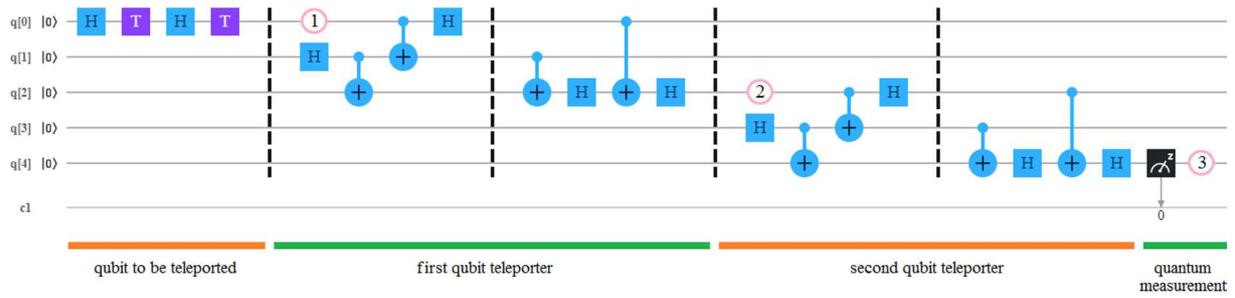

FIG. 13: The experiment of Fig. 11 implemented on IBM Q [39]. The qubit to be teleported is presented on point (1). After the first qubit teleporter procedure, the qubit disappears from Alice's side and appears in point (2) on the master satellite. After the second qubit teleporter procedure, the qubit disappears on the master satellite and appears in point (3) on Bob's side (on land).

(a) (b)

(c) (d)

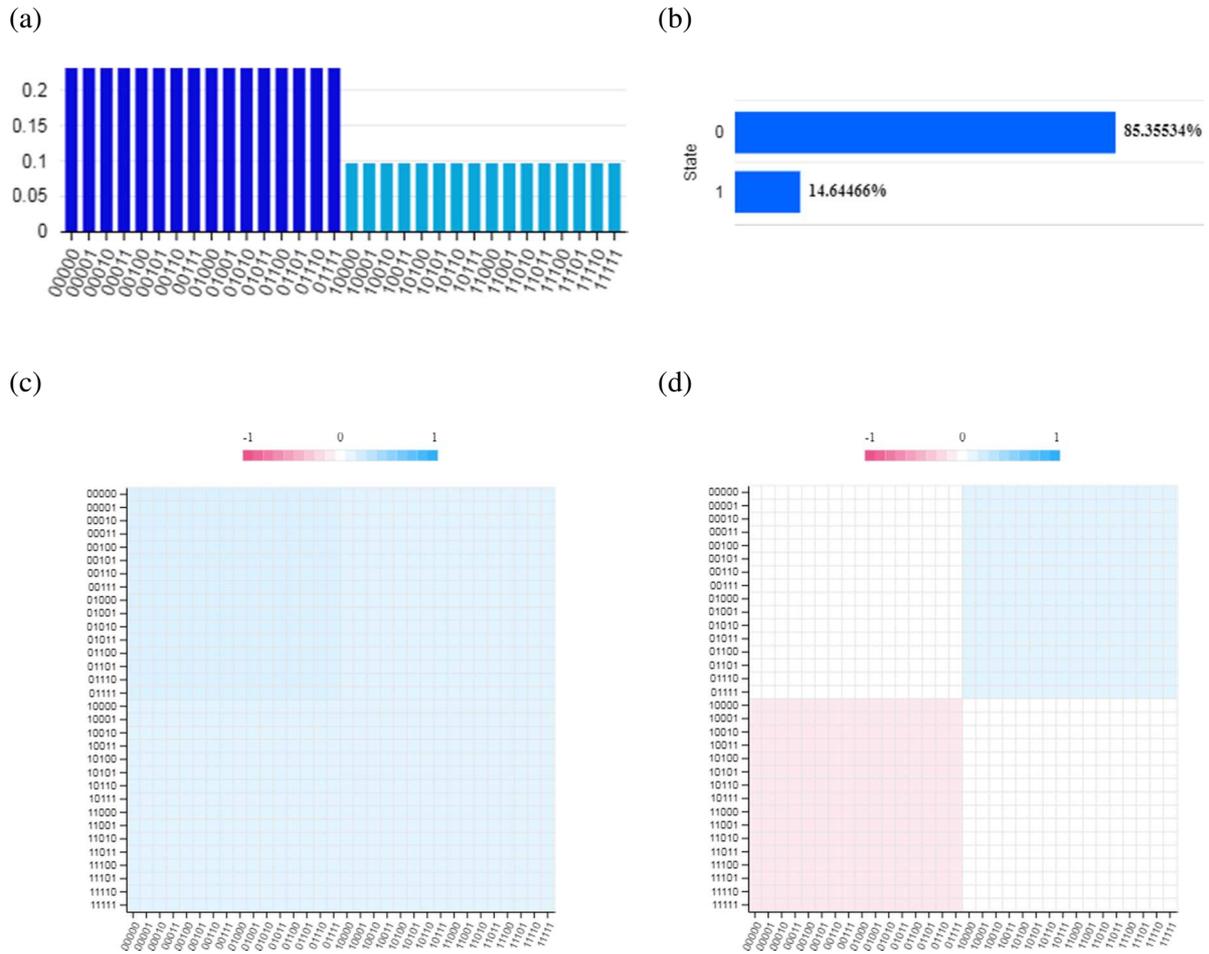

FIG. 14: Results of the teleportation of the state of Fig. 11 on IBM Q [39]: a) is the real part of the state (histogram) where only those non-zero are shown, b) the height of the bar is the complex modulus of the wavefunction, c) is the real part density matrix of the state, and d) is the imaginary part density matrix of the state.

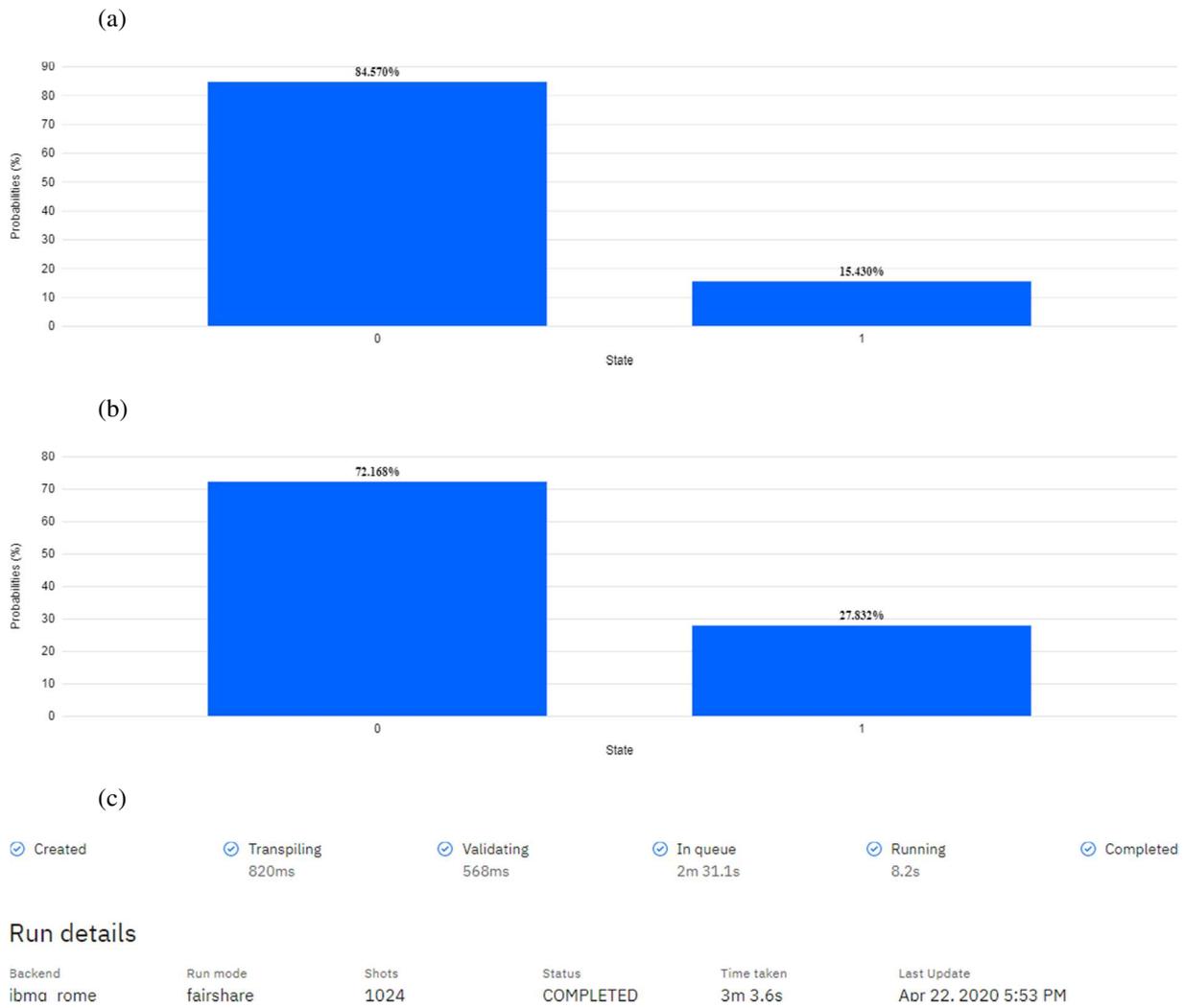

FIG. 15: Results of the teleportation of the state of Fig. 11 on IBM Q [39]: a) Probabilities (outcomes) of IBM QPU simulator [39], b) Probabilities (outcomes) of IBM QPU [39] for ibmq_rome processor, and c) Run details of IBM QPU [39] for ibmq_rome processor.